\documentstyle[preprint,aps]{revtex}
\begin{document}
\draft
\title {Preferred Basis in a Measurement Process}
\author {Anu Venugopalan}
\address { School of Physical Sciences,
Jawaharlal Nehru University,
New Delhi-110067, INDIA}
\maketitle
\begin{abstract}
The effect of decoherence is analysed for a free  particle,  interacting
with an environment via a dissipative coupling. The interaction  between
the particle and the environment occurs by a coupling  of  the  position
operator of the particle with the environmental degrees of  freedom.  By
examining the exact solution of the density matrix  equation  one  finds
that the density matrix becomes completely  diagonal  in  momentum  with
time while the position space  density  matrix  remains  nonlocal.  This
establishes  the  momentum  basis  as  the  emergent  'preferred  basis'
selected by the environment which is contrary to the general expectation
that position should emerge as the preferred basis  since  the  coupling
with the environment is via the position coordinate.
\end{abstract}

The act of measurement for quantum systems has defied  understanding  as
this process involves a collapse of the  state  vector  to  one  of  the
eigenstates of the dynamical  operator  which  is  being  measured.  The
process of collapse is  nonunitary  \cite{1}  and  cannot  be  described
quantum  mechanically.  In  recent  years,  the  'decoherence'  approach
\cite{2,3,4,5,6} to the quantum  measurement  problem  has  successfully
tackled many conflicts between the predictions of  conventional  quantum
theory and  classical  perceptions.   This  approach  seems  to  provide
convincing explanations for the  emergence  of  'classicality'  from  an
underlying quantum substrate. In this approach, the measuring apparatus,
which  is  often  macroscopic,  is  never  isolated  and  is  constantly
interacting  with  a  large  environment.  The  physical  system,  which
comprises of the  quantum  system,  the  measuring  apparatus,  and  the
environment, has a large number of independent parameters or degrees  of
freedom. However, we are often interested in  only  a  small  number  of
these degrees of freedom to  describe  the  outcome  of  a  measurement.
Decoherence is a consequence of 'ignoring' large numbers of  degrees  of
freedom. More technically, if we are descriing the  interaction  between
the apparatus and the environment by  a  density  matrix  \cite{6},  and
trace over the environment  degrees  of  freedom,  the  reduced  density
matrix of the apparatus is driven  diagonal  as  a  consequence  of  the
environmental influence.   The  pure  state  density  matrix  thus  gets
reduced to a mixture. This amounts to saying that superpositions  vanish
and the  density  matrix  can  be  interpreted  in  terms  of  classical
probabilities. However, even in this approach the question of the  basis
in which the density  matrix  becomes  diagonal,  i.e.,  the  'preferred
basis' \cite{6} in which superpositions vanish, is not quite understood.
It seems that regardless of the initial conditon, the environment always
selects a special set of states in which the density matrix is  diagonal
and hence classically interpretable \cite{6}. It is  important  to  know
what these preferred bases are  for  specific  systems  since  they  are
directly related to the emergent 'classicality' of  macroscopic  systems
as a  consequence  of  decoherence.  It  seems  plausible  that  such  a
preferred basis would be decided by the system operator which is coupled
to the environmental  degrees  of  freedom.  For  example,  for  a  free
particle if the position operator  is  involved  in  coupling  with  the
environmental degrees of freedom, one expects that  the  density  matrix
would be driven diagonal in the position space. It  has  been  shown  by
Zurek \cite{6} that the coherence  between  two  Gaussian  wave  packets
separated in space by $\Delta x$ is  lost  on  a  time  scale  which  is
typically

\begin{equation}
\theta=\tau \Big[{\hbar \over{\Delta x \sqrt{4 m k_{B}T}}}\Bigr]^2\label{1}
\end{equation}
where $m$ is the mass of the particle, $k_{B}$  is Boltzmann's constant,
$T$ is the temperature  of  the  heat  bath  and  $\gamma^{-1}$  is  the
characteristic relaxation time of the system. For classical systems  and
standard macroscopic separations $\Delta x$, the ratio $\theta  /  \tau$
can be as small as $10^{-40}$  \cite{6},  suggesting  that  the  density
matrix becomes diagonal in position space almost instantaneously, making
position the 'preferred basis'. Here, however, we show that for  a  free
particle such is not the case. This implies, therefore, that to  observe
a dynamical variable, the system-apparatus coupling requires  a  careful
consideration.
\section {The Master Equation}

The problem of a quantum system interacting with an environment has been
studied in great detail by  many  authors  in  the  context  of  quantum
dissipative systems \cite{7,8,9,10,11,12} and  the  quantum  measurement
problem  \cite{2,3,4,5,6,13,14}.  Here  we  employ  the  method  studied
extensively by Caldeira and Leggett \cite{9}  and  others  \cite{10}  to
study a free particle coupled to a collecton  of  harmonic  oscillators,
which constitutes the environment.  The  interaction  between  the  free
particle and the  environment  is  linear  via  a  coordinate-coordinate
coupling \cite{8,10}. The total Hamiltonian for the composite system can
be written as
\begin{equation}
H = {P^2\over{2m}} + \sum_j \left( {p_j^2\over{2m_j}}
+ {m_j \omega_j^{2} \over 2 }\Bigl[ x_{j} - {c_{j}Q \over m_{j}\omega_j^2}
 \Bigr]^2 \right).\label{2}
\end{equation}
Here, $P$, $Q$ are the momentum and position  coordinates  of  the  free
particle, and  $p_{j}$   and  $x_{j}$  ,   those  of  the  jth  harmonic
oscillator.  $c_{j}$s  are  coupling  strengths  and  $w_{j}$s  are  the
frequencies  of  the oscillators. The Hamiltonian of \ref{2},  is  known
as the independent oscillator  model.  More   frequently   seen  in  the
literature is the 'linear coupling' model  where  the   Hamiltonian   is
without  the 'counter term' ( last term in in the  summation)  and   the
coupling is represented by adding a term of the  form  $Q\sum_{j}  c_{j}
x$.  As pointed  out  by  Ford et al  \cite{8},  the  'linear  coupling'
Hamiltonian is unphysical and corresponds  to  a   'passive'  heat  bath
and is not invariant under translations.  All  information   about   the
harmonic  oscillator  heat  bath  which is required for the  description
of the particle via  a  reduced  density  matrix  is  contained  in  the
spectral density  function \cite{8,9} and the initial temperature of the
bath.  Using  the  Feynman-  Vernon   influence   functional   technique
\cite{11}, Caldeira and Leggett have shown that for  a   white   (ohmic)
noise spectrum  in  the   high  temperature  limit,  one  can  write  an
equation  of  motion  for  the reduced density  matrix  $\rho$   of  the
free particle. In the position representation this equation can  be  can
be written \cite{6,12} as
\begin{eqnarray}
{\partial{\rho_R(x,y,t)}\over\partial t)} &=& \Bigl[{-\hbar\over 2im}
\{ {\partial^2\over\partial{x^2}}
- {\partial^2\over\partial{y^2}} \}
- \gamma(x-y)\{ {\partial\over\partial x}
-{\partial\over\partial y} \}
- {D\over{4\hbar^2}}(x-y)^2 \Bigr] \rho_R(x,y,t),\label{3}
\end{eqnarray}
where $m$ is the mass of the particle,  $\hbar$  is  Planck's  constant,
$\gamma$ is the Langevin fricton  coefficient  and  $D$  has  the  usual
interpretation of the diffusion constant. $\gamma$ and $D$  are  related
to the parameters of the Hamiltonian \ref{2}.  For  a  high  temperature
thermal bath, $D=2m\gamma k_B T$. \ref{3} has been  used  to  study  the
dynamics of systems like the free particle and the  harmonic  oscillator
in interaction with a heat bath  \cite{6,7,8,9,10,11,12,13,14}.  In  the
following we will be looking at the exact solution  of  \ref{3}  in  the
position and momentum representations.

\section {Preferred Basis}

Consider the  exact  solutions  of  \ref{3}  derived  earlier  by  Kumar
\cite{12} for an initial Gaussian wave packet
\begin{equation}
\psi(x,0)={1\over {(\sigma \sqrt{\pi})^{1/2}}} \exp(-x^2/2\sigma^2)\label{4}
\end{equation}
where  $\sigma$  is  the  width  of  the  wave  packet.  The   solutions
\cite{12,16} in wave vector and position representations in terms of the
changed coordinates $Q = u-v$; $q = (u + v)/2$ in wave vector space  and
$R = (x + y)/2$; $r = (x-y)$ in position space, are
\begin{eqnarray}
\rho_d (Q,q,t) &=& 2 \sqrt{\pi\over {N(\tau)}} \exp\Biggl[ {-1\over
 N(\tau)} \Big[ q
+ {i\hbar Q\over 2\sigma^2m\gamma} e^{-\tau} (1-e^{-\tau})
- {iQD\over 4\hbar\gamma^2m}(1-e^{-\tau})^2 \Bigr]^2 \nonumber\\
&-& \Big[{\hbar^2\over 4\sigma^2m^2\gamma^2}(1-e^{-\tau})^2 +
{\sigma^2\over 4}
+ {D\over 2m^2\gamma^3}
(2\tau-3+4e^{-\tau}-e^{-2\tau})\Bigr] Q^2\Biggr] \label{5}
\end{eqnarray}
where $\tau = \gamma t$ and
\begin{equation}
N(\tau) \equiv (D/2\hbar^2\gamma)(1-e^{-2\tau})+(1/\sigma^2) e^{-2\tau},
\label{6}
\end{equation}
and
\begin{eqnarray}
\rho_d (R,r,t) &=& 2 \sqrt{\pi\over {N(\tau)}} \exp\Biggl[ -\Big[ {1\over
4\sigma^2} e^{-2\tau}
+ {D\over 8\hbar^2\gamma} (1-e^{-2\tau}) \Bigr] - {1\over M(\tau)} \Big[ R
\nonumber\\
&-& {i\hbar r\over 2\sigma^2m\gamma} e^{-\tau}(1-e^{-\tau})
- {iDr\over 4\hbar\gamma^2m} (1-e^{-\tau})^2 \Bigr]^2 \Biggr]
\label{7}
\end{eqnarray}
where
\begin{eqnarray}
M(\tau) &\equiv& \sigma^2 +{\hbar^2\over \sigma^2 m^2\gamma^2}(1 -
e^{-\tau})^2
+ {D\over 2m^2\gamma^3} ( 2\tau - 3 + 4e^{-\tau} - e^{-2\tau}).\label{8}
\end{eqnarray}
The time dependence of the above two solutions can be  studied  for  two
regimes:  $\tau  >>  1$,  i.e.,  for  times   much   larger   than   the
characteristic relaxation time $\gamma^{-1}$, and $\tau << 1$, i.e., for
times much smaller than $\gamma^{-1}$. The first regime $(\tau >> 1)$ is
of greater importance in all real life systems.
\subsection{$ \tau >>1 $}
It  is  clear  from  the  form  of  the   solution   \ref{5}   that   as
$\tau\rightarrow\infty$, the off-diagonal elements of the density matrix
in the momentum representation vanish. This  decay  is  exponential  and
occurs in a time

\begin{equation}
 t_d = {m^2\gamma^2\over D Q^2} . \label{9}
\end{equation}
The density matrix \ref{5} becomes completely diagonal at long times and
assumes the form
\begin{equation}
\rho(0,u,t) = 2 \sqrt{\pi\over N(\tau)} \exp \left\{ {-u^2\over N(\tau)}
\right\}, \label{10}
\end{equation}
making momentum the obvious choice for the preferred basis. \ref{10}  is
the  classical  Maxwell  distribution  that  one  would  obtain  from  a
classical Fokker-Planck equation. We would like to point out  here  that
the expression \ref{9} for $t_d$  is valid for the regime  $\tau  >>  1$
which implies a time scale much greater than the relaxation time of  the
system $\gamma^{-1}$.  Moreover \ref{3} is valid for a high  temperature
heat bath, where $\hbar\gamma /k_B T << 1$ \cite{9}. From the expression
\ref{9} one can see that  the  decoherence  time  $t_d$  decreases  with
increasing $T$ and $\gamma$. On the other hand the distribution function
\ref{7} in  the  position  representation  does  not  become  completely
diagonal as $\tau\rightarrow\infty$, but assumes the form
\begin{eqnarray}
\rho_d (R,r,t) &=& 2 \sqrt{\pi\over M(\tau)} \exp\biggl[ -
{D\over 8\hbar^2\gamma} r^2
- {1\over M(\tau)} \Bigl[ R + {iDr\over 4\hbar\gamma^2m}
 (1-e^{-\tau})^2 \Bigr]^2 \biggr], \label{11}
\end{eqnarray}
with
\begin{equation}
M(\tau) \simeq {D\tau\over m^2\gamma^3}. \label{12}
\end{equation}
The density matrix is  obviously  non-diagonal  in  the  position  space
representation which is a consequence  of  the  fact  that  \ref{5}  and
\ref{7} are related by Fourier transforms. The asymptotic width for  the
distribution in variable $r$ is $D/8\hbar^2\gamma$, which for a  thermal
bath is $\pi/2\lambda_d^2$, since
\begin{equation}
\hbar^2\gamma/D = \lambda_{d}^{2}/4\pi, \label{13}
\end{equation}
where $\lambda_d$  is the thermal de Broglie wavelength of the  particle
($h/\sqrt{2m\pi  k_B  T}$).  One  can  see  that  if   the   extent   of
'off-diagonality' is much greater than $\lambda_d$, the magnitude of the
off-diagonal elements, which is weighted by  $e^{-r^2\pi/2\lambda_d^2}$,
is very small and the denity matrix in position space can be  considered
nearly diagonal. In principle,  however,  it  remains  nonlocal  to  the
extent of the deBroglie wavelength, $\lambda_d$. It  is  interesting  to
see that for the initial condition considered by Zurek  \cite{6},  where
the initial position-space density matrix contains  four  well-separated
peaks ($\Delta x >> \sigma$), an exact solution of  \ref{3}  shows  that
for the peaks which are along the diagonal for  which  $R\simeq\pm\Delta
x; r \simeq 0$, the major contribution to the density matrix is from the
diagonal elements, which are peaked around $\pm\Delta x$.  However,  for
the  peaks  along  the  off-diagonal,  for  which  $R\simeq  0;  r\simeq
\pm\Delta x$ the factor $\exp (-r^2\pi/2\lambda_d^2)$ (see \ref{11})  is
now $\exp (-\Delta x^2\pi/2\lambda_d^2)$. One can see that for $\Delta x
>> \lambda_d$, this factor is very small and hence the elements  of  the
density  matrix  corresponding  to  these  two  off-diagonal  peaks  are
negligible  in  magnitude.  Thus,  position  seems  to  emerge   as   an
approximate preferred basis, which works well only when one  is  probing
length scales which are much larger than $\lambda_d$.
\subsection{$ \tau << 1 $}

In this regime, one is probing the system at time scales which are  much
smaller than the  characteristic  relaxation  time  $\gamma^{-1}$.  This
regime is not very realistic since one  is  usually  interested  in  the
state of the system long after it has been left in an environment. If we
retain terms only upto first order in $\tau$ in \ref{5} and \ref{7}, the
momentum and position space density matrices assume the forms
\begin{eqnarray}
\rho_d (Q,q,t) &=& 2\sqrt{\pi\over N(\tau)} \exp \Biggl[{-1\over N(\tau)}
\Bigl[q
+ {i\hbar Q\over 2\sigma^2m\gamma}\tau(1-\tau)
- {iQD\over 4\hbar\gamma^2 m} \tau^2\Bigr]^2 \nonumber\\
&-& \Bigl[{\hbar^2\over 4\sigma^2m^2\gamma^2}\tau^2
+ {\sigma^2\over 4}\Bigr] Q^2\Biggr],\label{14}
\end{eqnarray}
where
\begin{equation}
N(\tau) \simeq (D\tau /\hbar^2\gamma) + (1/\sigma^2) (1-2\tau ),\label{15}
\end{equation}
and
\begin{eqnarray}
\rho_d (R,r,t) &=& 2\sqrt{\pi\over M(\tau)} \exp \Biggl[-\Bigl[{1\over
4\sigma^2
}(1-2\tau)
+ {D\tau\over 4\hbar^2\gamma}\Bigr]r^2 - {1\over M(\tau)}\Bigl[ R \nonumber\\
&-& {i\hbar r\over 2\sigma^2 m\gamma}\tau (1-\tau) + {iDr\over
4m\gamma^2\hbar}
\tau^2\Bigr]^2 \Biggr], \label{16}
\end{eqnarray}
where
\begin{equation}
M(\tau) \simeq \sigma^2 + {\hbar^2\tau^2\over \sigma^2m^2\gamma^2}. \label{17}
\end{equation}
It is clear from \ref{14} and \ref{16} that the density matrices in both
representations remain nondiagonal  and  neither  show  any  exponential
decay with time for the off-diagonal elements. For the position  density
matrix \ref{16}, the leading order decaying term is $\sim \exp {\left( -
{1\over 2}\{ D/2\hbar^2\gamma - 1/\sigma^2 \} r^2 \tau\right)  }$.  Note
that $D r^2\tau\sim 4\pi r^2\tau/\lambda_d^2$ and the  'decay'  time  is
similar to \ref{1} obtained by Zurek \cite{6}. However,  since  this  is
valid only for $\tau << 1$, it  cannot  be  interpreted  as  a  complete
exponential decay of the off- diagonal elements in  position  space.  In
this time regime, thus, both the position and the momentum space density
matrices remain highly nonlocal and one cannot  talk  in  terms  of  any
emergent preferred basis.

 From the above analysis it is clear that in the  $\tau\rightarrow\infty$
regime, which is a significant regime for  realistic  systems,  momentum
emerges as the basis selected by the  environment.  The  position  space
density  matrix  remains  nonlocal,  the  extent  of  nonlocality  being
$\lambda_d$ .

To summarize, we have clarified the roles of position and momentum for a
free particle which is dissipatively coupled to a heat bath. The  nature
of the emergent preferred basis is expected to depend on the form of the
system-environment   coupling.   We   use   a   Hamiltonian    with    a
coordinate-coordinate coupling between the system and  the  environment,
but  it  is  unitarily   equivalent   to   other   types   of   coupling
(coordinate-momentum {\it etc.}) \cite{8}. The preferred basis need  not
necessarily be  that  of  the  system  variable  which  couples  to  the
envirnmental degrees of freedom. For the coordinate-coordinate coupling,
we have shown that the momentum basis clearly is the emergent  preferred
basis.  This is contrary to the general expectation that position should
emerge as the preferred basis.

The author acknowledges finanacial support from  the  University  Grants
Commision, India and wishes to thank Deepak Kumar and  R.  Ghosh  for  a
critical reading of the  manuscript  and  encouraging  discussions.  The
author also wishes to thank T. Qureshi for many useful discussions.  The
author thanks the Referee for bringing to our notice refernces 8 and 10.

\end {document}